\documentstyle[prb,twocolumn,aps,rotate,epsf,floats]{revtex}

\newcommand{\mCu}{{\mbox{\tiny Cu}}}
\newcommand{\mGe}{{\mbox{\tiny Ge}}}
\newcommand{\mOa}{{\mbox{\tiny O2a}}}
\newcommand{\mOb}{{\mbox{\tiny O2b}}}
\newcommand{\mO}{{\mbox{\tiny O2}}}
\newcommand{\loc}{{\mbox{\tiny loc}}}

\begin{document}
\setlength{\unitlength}{1cm}
\renewcommand{\arraystretch}{1.4}

\title{The microscopic spin-phonon coupling constants in CuGeO$_3$}

\author{Ralph Werner and Claudius Gros}
\address{Institut f\"ur
 Physik, Universit\"at Dortmund, D-44221 Dortmund, Germany.}
\author{Markus Braden}
\address{Forschungszentrum Karlsruhe, Institut f\"ur
   Nukleare Festk\"orperphysik, Postfach 3640, D-76021 Karlsruhe, Germany\\and
   Laboratoire L\'eon Brillouin (CEA-CNRS), Centre d'\'Etudes
   Nucl\'eaires de Saclay, F-91191 Gif-sur-Yvette C\'edex, France.}

\date{\today}

\maketitle

\centerline{Preprint. Typeset using REV\TeX}

\begin{abstract}
Using random phase approximation results, mean-field theory, and
refined data for the polarization vectors we determine the coupling
constants of the four Peierls-active phonon modes to the spin
chains. We then derive the values of the coupling of the spin system
to the linear ionic displacements, the bond lengths and the angles
between bonds. Our values are consistent with microscopic theories and
various experimental results. We discuss the applicability of static
approaches to the spin-phonon coupling. The $c$-axis anomaly of the
thermal expansion is explained. We give the values of the coupling
constants in an effective one-dimensional Hamiltonian.
\end{abstract}


\section{Introduction}
With the discovery of the first inorganic spin-Peierls
compound CuGeO$_3$,\cite{Hase93CGO} it has become possible to
investigate the physics of the spin-Peierls transition in
quasi-one-dimensional spin chains with high precision. As a consequence, it
has attracted much attention both in experiment and theory. Yet, for a
long time the absence of a soft phonon mode
\cite{Hirota95CGO,Lorenzo94CGO} has been puzzling, since the behavior
was believed not to be consistent with the standard approach to
spin-Peierls transitions by Cross and Fisher.\cite{CF79} The
frequencies of the Peierls-active phonon modes being of the order of
the magnetic exchange,\cite{Braden98CGO} it has been argued
that the Cross and Fisher approach is not applicable because of the
nonadiabacity of the phonons.\cite{Uhrig98SP} Only recently we were
able to show\cite{Gros98CGO} that the random-phase-approximation (RPA)
approach by Cross and Fisher indeed is consistent with the hardening
of the Peierls-active phonon modes.\cite{Braden98CGO}

It is now tempting to combine the RPA results with the
detailed data on the phonons acquired by Braden {\it et~al.}
\cite{Braden98CGO} Treating the lattice with the standard harmonic
theory and including the spin-phonon coupling mean-field like, we
calculate the microscopic coupling constants between the lattice and
the spin chains. It is then possible to predict the effect of
structural changes on the spin system, namely the antiferromagnetic
exchange $J$. The latter has been subject to various experimental
\cite{Lorenz96CGO,Takahashi95CGO,Nishi95CGO} and
theoretical studies.\cite{Riera95CGO,Castilla95CGO,Chitra95DMRG,Geertsma96CGO,Braden96CGO,Brenig97CGO,Buechner98CGO}

The microscopic three-dimensional (3D) Hamiltonian we have to consider
consists of three parts:
\begin{equation}
H = H_s + H_{p} + H_{sp}.
\end{equation}
The Heisenberg spin Hamiltonian
\begin{equation}\label{Hs}
H_s=J\ \sum_{\bf l}\ {\bf S}_{\bf l}\cdot{\bf S}_{{\bf l}+\hat{z}}
    + J_2\ \sum_{\bf l}\ {\bf S}_{\bf l}\cdot{\bf S}_{{\bf l}+2\hat{z}}
\end{equation}
with the exchange integrals $J$ and $J_2$ between nearest-neighbor
(NN) and next-nearest-neighbor (NNN) Cu $d$ orbitals, respectively. 

Further we distinguish the phonon part 
\begin{equation}\label{Hph}
H_{p}=\sum_{{\bf n}\atop\nu,\alpha}
                \frac{\left(p^\alpha_{{\bf n},\nu}\right)^2}{2 m_\nu}
        + \sum_{{\bf n},{\bf n}'\atop\nu,\nu',\alpha,\alpha'}
        \Phi_{{\bf n},{\bf n}',\nu,\nu'}^{\alpha,\alpha'}\,
        r_{{\bf n},\nu}^\alpha\,r_{{\bf n}',\nu'}^{\alpha'},
\end{equation}
describing the lattice vibrations in the harmonic approximation, where
${\bf r}_{{\bf n},\nu} 
=(r_{{\bf n},\nu}^x, r_{{\bf n},\nu}^y, r_{{\bf n},\nu}^z)$
are the deviations from the ionic equilibrium positions. 

Finally the spin-phonon coupling term reads
\begin{equation}\label{Hsp}
H_{sp}=\sum_{\bf l}
        \Delta J_{{\bf l},{\bf l}+\hat{z}}
        \ {\bf S}_{\bf l}\cdot{\bf S}_{{\bf l}+\hat{z}}+
\sum_{\bf l}
        \Delta J_{{\bf l},{\bf l}+2\hat{z}}
        \ {\bf S}_{\bf l}\cdot{\bf S}_{{\bf l}+2\hat{z}}.
\end{equation}
The energy scale $\Delta J_{{\bf l},{\bf l}+\hat{z}}$
is a function of the variation of the magnetic exchange integral with
the atomic displacements $g^\alpha_{{\bf l},\nu} = \partial J /
\partial r^\alpha_{{\bf l},\nu}$ to be discussed in
Sec.~\ref{sectionS-P-C}. The NNN term $\Delta J_{{\bf l},{\bf
l}+2\hat{z}}$ is a function of $\partial J_2 / \partial r^\alpha_{{\bf
l},\nu}$.

\begin{figure}[tb]
\epsfysize=0.5\textwidth
\centerline{\rotate[r]{\epsffile{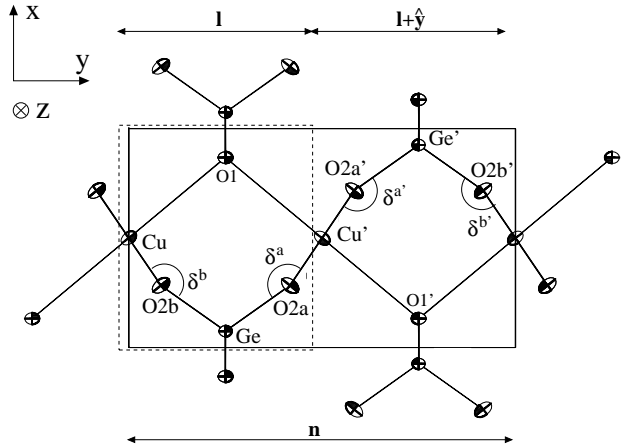}}}
\centerline{\parbox{0.48\textwidth}{\caption{\label{unitcell}
Projection of the unit cell of CuGeO$_3$ on the $x$-$y$ plane. The
oxygen atoms are distinguished into O1, O2a, and O2b, the atoms of
the second formula unit are labeled with a prime. Each unit cell
contains two Cu chains in $z$ direction (positive $z$ direction is
into the plane). The broken lines show the reduced unit cell
introduced in Sec.\ \protect\ref{sectionS-P-C}. ${\bf n}$ is the
index for the whole cells, ${\bf l}$ indexes the reduced cells. 
}}}
\end{figure}

The indices used are ${\bf n}=(n_x,n_y,n_z)\in {\cal Z}^3$ running
over all unit cells of the three-dimensional crystal, the Cu-site index
${\bf l}=(l_x,l_y,l_z)\in {\cal Z}^3$ (two Cu sites per unit cell),
and the unit vectors
$\hat{x}=(1,0,0)$, $\hat{y}=(0,1,0)$, and  $\hat{z}=(0,0,1)$ to
nearest-neighbor unit cells in the corresponding direction.
The index $\nu$ labels the 10 atoms within a unit cell as shown in
Fig.~\ref{unitcell} and $\alpha\in\{x,y,z\}$ is the vectorial
component of the indexed quantity in the respective three-dimensional
space. 

In Sec.\ \ref{sectionTplus} we briefly summarize the
diagonalization of the phonon Hamiltonian (\ref{Hph}) followed by the
discussion of the symmetry of the four Peierls-active phonon modes,
including refined data for their polarization vectors. Using these
symmetries we transform in Sec.\ \ref{sectionS-P-C} the microscopic
spin-phonon coupling Hamiltonian (\ref{Hsp}) to normal coordinates in
reciprocal space. This procedure yields relations between the different
linear, angular, and normal mode coupling constants. Using RPA results
and mean-field theory in Sec.\ \ref{sectionnormal} we obtain
numerical values for the normal-mode coupling constants which then can
be converted to the real-space coupling constants. The resulting
dependence of the magnetic exchange on static distortion of the
lattice is discussed in Sec.\ \ref{sectionstatic} and compared with
values from the literature. Finally we derive an effective
one-dimensional model to give coupling constants consistent with 
frequently applied theoretical approaches. The consistency of the
different results gives a \`a posteriori justification of the
mean-field approach.


\section{Peierls-active phonon modes}\label{sectionTplus}

In the standard treatment of harmonic lattice dynamics, the initial
problem of $3\cdot N \cdot N_{\rm ion}$ degrees of freedom ($N$ number
of unit cells, $N_{\rm ion}$ number of ions in the unit cell) is
transformed into reciprocal space by a Fourier transformation, where
$N$ wave vectors fulfill the periodic boundary
condition.\cite{BornHuang} For any fixed wave vector one obtains a
$3\cdot N_{\rm ion}$-dimensional problem which may be diagonalized,
resulting in a set of $3\cdot N_{\rm ion}$ eigenmodes labeled by
$\lambda\in\{1,\ldots,3N_{\rm ion}\}$. For that purpose, the
displacement and momentum operators are decomposed into eigenmode
contributions introducing normal coordinates $Q$ and conjugated
momenta $P$:
\begin{eqnarray}
{\bf r}_{{\bf n},\nu}&=
        &\frac{1}{\sqrt{N}}\sum_{\bf q} e^{i{\bf q}{\bf R}_{{\bf n}}}
        \ \sum_\lambda\frac{{\bf e}_\nu(\lambda,{\bf q})}{\sqrt{m_\nu}}
        \ Q_{\lambda,{\bf q}} \label{xdecomp},
\\
{\bf p}_{{\bf n},\nu}&=
        &\frac{1}{\sqrt{N}}\sum_{\bf q} e^{i{\bf q}{\bf R}_{{\bf n}}}
        \ \sum_\lambda {\bf e}_\nu(\lambda,{\bf q})\sqrt{m_\nu}
        \ P_{\lambda,{\bf q}} \label{pdecomp}.
\end{eqnarray}
The vectors ${\bf R}_{{\bf n}}$ designate the coordinates of the unit-cell
origins. $m_\nu$ is the mass of the $\nu$th atom and ${\bf e}_\nu$ are
polarization vectors. Note that we use a nonstandard definition for
${\bf R}_{{\bf n}}$ and ${\bf e}_\nu(\lambda,{\bf q})$ which will
simplify the interpretation of the polarization vectors at high
symmetry points in the Brillouin zone. Further transformation to boson
creation and annihilation operator representation via
\begin{eqnarray}
\label{Qtob}
Q_{\lambda,{\bf q}}&=&\sqrt{\frac{\hbar}{2\Omega_{\lambda,{\bf q}}}}
        \left(b^\dagger_{\lambda,-{\bf q}}+b^{}_{\lambda,{\bf q}}\right),
\\
\label{Ptob}
P_{\lambda,{\bf q}}&=&i\sqrt{\frac{\hbar\Omega_{\lambda,{\bf q}}}{2}}
        \left(b^\dagger_{\lambda,-{\bf q}}-b^{}_{\lambda,{\bf q}}\right),
\end{eqnarray}
yields the Hamiltonian usually used in the theoretical treatment of
the lattice vibrations
\begin{equation}\label{Hpbose}
H_p=\sum_{{\bf q},\lambda}\hbar\Omega_{\lambda,{\bf q}}
\left(b^\dagger_{\lambda,{\bf q}}b^{}_{\lambda,{\bf
q}}+\frac{1}{2}\right).
\end{equation}

With the experimentally determined phonon modes $\Omega_{\lambda,{\bf
q}}$ and shell-model calculations, it is possible to determine the
components of the polarization vectors ${\bf e}_\nu(\lambda,{\bf
q})$.\cite{Braden98CGO} At the wave vector of the Peierls instability
${\bf q}_0 = (\pi/a,0,\pi/c)$, four of the 30 modes correspond to the
irreducible representation with the symmetry of the lattice distortion
in the spin-Peierls phase, T$_2^+$ in the notation of Ref.\
\onlinecite{stokes}. $a=4.8$ {\AA}, $b=8.5$ {\AA}, and $c=2.9$ {\AA}
are the lengths of the unit cell in $x$-, $y$-, and $z$-direction,
respectively.  

\begin{figure}[tb]
\epsfxsize=0.48\textwidth
\centerline{\epsffile{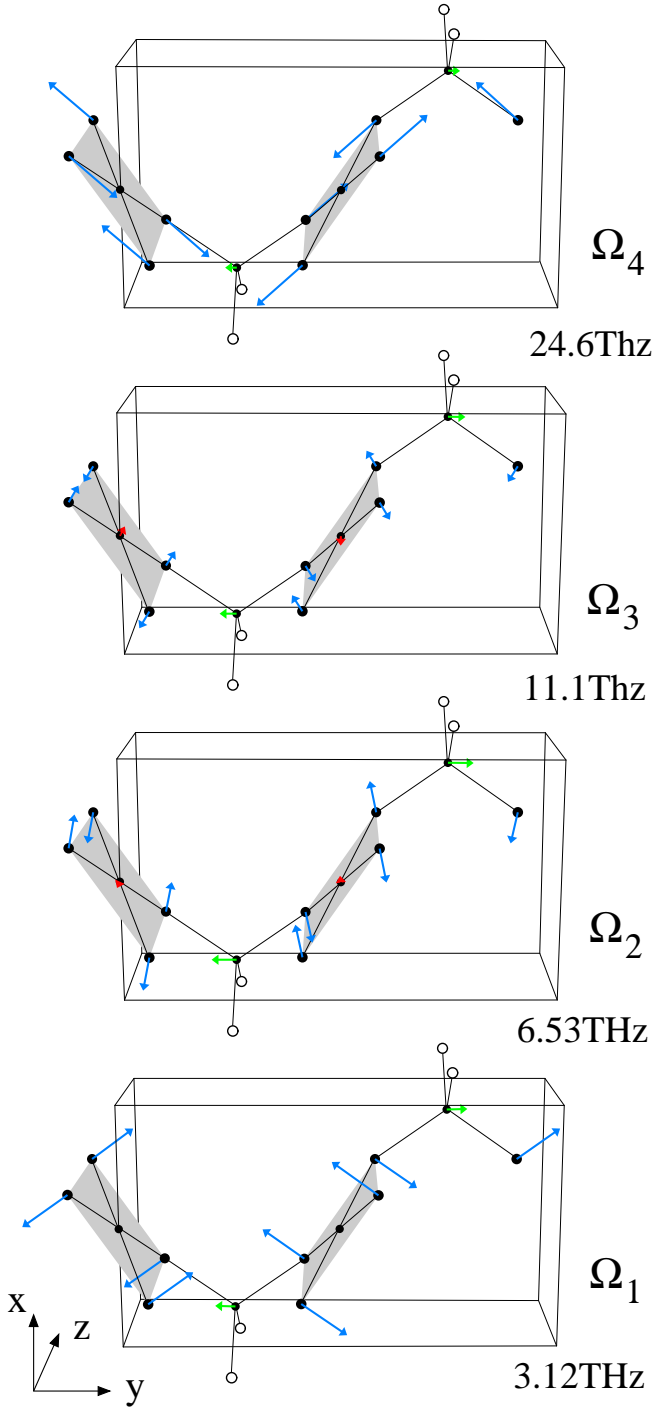}}
\centerline{\parbox{0.48\textwidth}{\caption{\label{T2Modes}
Geometry of the T$_2^+$ eigenmodes as given by the polarization
patterns in Table \protect\ref{tablefrequencies}. The shaded areas are
the CuO$_4$ plaquettes which form the Cu chains in $z$ direction. The
Cu atoms are in the center of each plaquette, the corners are formed
by O2 ions. The O1 atoms are represented by the open circles with the
Ge ions in between them. (Compare with the $x$-$y$ projection given in
Fig.\ \protect\ref{unitcell}.) Note that the O2 elongations are in the
$x$-$y$ plane, the (small) Cu displacements are along the $z$ Axis
while the Ge displacements are along $y$.
}}}
\end{figure}

Adapting the lattice-dynamical model presented in 
Ref.\ \onlinecite{Braden98CGO}, a special effort was made for the
description of the spin-Peierls relevant modes by the introduction of
additional force constants; details of the lattice-dynamical study
will be given elsewhere.\cite{Bradentbp} The T$_2^+$ modes are
characterized by displacements of the Cu ions along $c$, of the
O2 sites along $a$ and $b$, and of the Ge ions along $b$. 
The polarization pattern of the four modes as obtained by the
shell model are represented in Fig.~\ref{T2Modes} and given together
with their frequencies in Table \ref{tablefrequencies}. 
The highest T$_2^+$ mode corresponds to a Ge-O bond stretching
vibration thereby explaining its elevated frequency.
The three T$_2^+$ modes at lower energies posses a common element
which consists in the rotation of the O2-O2 edges of the
CuO$_4$ plaquettes in the $x$-$y$ plane around the $c$ axis. However,
only for the lowest mode does this twisting of the 
CuO$_2$ ribbons  describe the main character of the polarization
pattern. The modes at 11 and at 6.5 THz show, in addition,
a modulation of the lengths of the O2-O2 edges and a Cu shift
parallel $c$. For the 11 THz mode the displacements of the Cu ions
modulate the O2-Cu bond distance. The 6.5 THz mode is characterized by
a strong modulation of the Cu-O2-Cu bond angle (see
Fig.~\ref{T2Modes}) which is essential for the magnetic interaction.

\begin{table}[tb]
\begin{tabular}{lrrrr}
$\lambda$ &
\multicolumn{1}{c}{1} & \multicolumn{1}{c}{2} &
\multicolumn{1}{c}{3} & \multicolumn{1}{c}{4} \\\hline
\multicolumn{1}{c}{$\Omega_\lambda/(2\pi)$} &
         3.12 THz & 6.53 THz & 11.1 THz & 24.6 THz \\\hline
$u\,{\rm e}^z_\mCu(\lambda)/\sqrt{m_\mCu}$ &
        0.0095 & -0.4790 & 0.7412 & -0.0888 \\
$u\,{\rm e}^y_\mGe(\lambda)/\sqrt{m_\mGe}$ &
        -0.4330 & -0.5325 & -0.3698 & -0.2605\\
$u\,{\rm e}^x_\mOb(\lambda)/\sqrt{m_\mO}$ &
        -0.6212& 0.6581 & 0.3382 & -0.7932\\
$u\,{\rm e}^y_\mOb(\lambda)/\sqrt{m_\mO}$ &
        -0.8620 & 0.1339 & 0.2021 & 0.8723
\end{tabular}
\caption{Frequencies and polarization of the Peierls-ac\-tive
T$_2^+$ phonon modes at room temperature. The global prefactor
is given by $u^2=(8.26\pm0.02)\times 10^{-26}$ kg. The notation is 
${\rm e}^z_\nu(\lambda,{\bf q}_0)\equiv{\rm e}^z_\nu(\lambda)$.}
\label{tablefrequencies}
\end{table}

For later use we define the matrix ${\bf M}$ with the elements
given by ${\rm e}^\alpha_\nu(\lambda,{\bf q}_0)/\sqrt{m_\nu}$
extracted from Table \ref{tablefrequencies}:
\begin{equation}\label{eMatrix}
{\bf M}=\left(\begin{array}{rrrr}
0.03 & -1.67 &  2.58 & -0.31 \\
-1.51 & -1.85  & -1.29  & -0.91 \\
-2.16 & 2.29 & 1.18 & -2.76 \\
-3.00 & 0.47 & 0.70 & 3.04
\end{array}\right)\frac{10^{12}}{\sqrt{\rm kg}}.
\end{equation}
The static distortion in the spin-Peierls phase at $T=4$~K also has
been determined.\cite{Braden96CGO} We define a corresponding
four-dimensional vector:
\begin{equation}\label{rVector}
\langle {\bf r} \rangle_{T=4{\rm K}}=
\left(\begin{array}{c}
\langle r^z_\mCu \rangle \\
\langle r^y_\mGe \rangle \\
\langle r^x_\mOb \rangle \\
\langle r^y_\mOb \rangle \\
\end{array}\right)=10^{-2}
\left(\begin{array}{r}
0.57 \\ 0.08 \\ -0.95 \\ -0.65
\end{array}\right){\rm \AA}.
\end{equation}


\section{Spin-phonon coupling term}
\label{sectionS-P-C}

In the spin-phonon coupling term (\ref{Hsp}) we focus on the NN part
for reasons that will become obvious at the end of the section. We
include the relevant displacements of the ions directly involved in
the Cu-Cu superexchange path determining $J$ and only those coupling
constants where the ions actually show displacements in the
Peierls-active modes. The apex O1 atoms are not displaced by those
modes at the appropriate wave vector ${\bf q}_0$. We have to consider
two copper ions in adjacent unit cells along the $c$ direction, two 
germanium sites, and two oxygen atoms  surrounding a Cu-Cu bond.
The notation introduced is shown in Fig.~\ref{unitcell}, the two
O2 oxygen atoms per formula unit are denoted O2a and
O2b. There are two formula units per unit cell which we
distinguish by a prime.

The relevant coupling constants for the linear atomic elongations are
shown in Table \ref{tablecouplings}.
The effective spin-phonon coupling Hamiltonian is 
\begin{eqnarray}\label{Hspeff}
H^{\rm NN}_{sp}\!&=&\!\sum_{\bf n}\!\Bigg[
g^z_{\mCu}\! \left(r^z_{\mCu,{\bf n}}\! - r^z_{\mCu,{\bf
n}+\hat{z}}\right)
- g^y_{\mGe}\! \left(r^y_{\mGe',{\bf n}-\hat{y}}\! - r^y_{{\mGe},{\bf
n}}\right)
                                                \nonumber\\&&\qquad
        -\; g^x_{\mOb}\! \left(r^x_{\mOb',{\bf n}-\hat{y}}
                - r^x_{{\mOb},{\bf n}}\right)
                                                \nonumber\\&&\qquad
        -\; g^y_{\mOb}\! \left(r^y_{\mOb',{\bf n}-\hat{y}}
                - r^y_{{\mOb},{\bf n}}\right)
\!\Bigg]\ {\bf S}_{\bf n}\cdot{\bf S}_{{\bf n}+\hat{z}}
\nonumber\\&+&
\sum_{\bf n}\Bigg[
g^z_{\mCu}\! \left(r^z_{\mCu',{\bf n}} - r^z_{\mCu',{\bf
n}+\hat{z}}\right)
- g^y_{\mGe\!} \left(r^y_{\mGe,{\bf n}} - r^y_{{\mGe'},{\bf n}}\right)
                                                \nonumber\\&&\qquad
    +\; g^x_{\mOa}\! \left(r^x_{\mOa,{\bf n}} - r^x_{{\mOa'},{\bf
n}}\right)
                                                \nonumber\\&&\qquad
    +\; g^y_{\mOa}\! \left(r^y_{\mOa,{\bf n}} - r^y_{{\mOa'},{\bf
n}}\right)
\!\Bigg]\ {\bf S}'_{\bf n}\cdot{\bf S}'_{{\bf n}+\hat{z}}\,.
\end{eqnarray}
The two sums correspond to
the two Cu chains running through each unit cell. 
\begin{table}[b]
\begin{tabular}{lclclcl}
$g^z_{\mCu}$ &=& $\frac{\partial J}{\partial r^z_{\mCu,{\bf n}}}$ &=&
        $-\frac{\partial J}{\partial r^z_{\mCu,{\bf n}+\hat{z}}}$ &=&
        $\ \ \,\frac{\partial J'}{\partial r^z_{\mCu',{\bf n}}}$
\\[2ex]
$g^y_{\mGe}$ &=& $\frac{\partial J}{\partial r^y_{\mGe,{\bf n}}}$ &=&
        $-\frac{\partial J}{\partial r^y_{\mGe',{\bf n}-\hat{y}}}$ &=&
        $-\frac{\partial J'}{\partial r^y_{\mGe,{\bf n}}}$
\\[2ex]
$g^{x,y}_{\mOa}$ &=&
        $\frac{\partial J'}{\partial r^{x,y}_{\mOa,{\bf n}}}$ &=&
        $-\frac{\partial J'}{\partial r^{x,x}_{{\mOa'},{\bf n}}}$ &=&
        $-\frac{\partial J'}{\partial r^{x,y}_{\mOa',{\bf n}-\hat{y}}}$
\\[2ex]
$g^{x,y}_{\mOb}$ &=&
        $\frac{\partial J}{\partial r^{x,y}_{\mOb,{\bf n}}}$ &=&
        $-\frac{\partial J}{\partial r^{x,y}_{{\mOb'},{\bf n}}}$ &=&
        $-\frac{\partial J}{\partial r^{x,y}_{\mOb',{\bf n}-\hat{y}}}$
\end{tabular}
\caption{Definition of the coupling constants for linear atomic
elongations. The two Cu chains running through each unit cell are
distinguished by a prime (see Fig.~\protect\ref{unitcell}). $J'$ is
the magnetic coupling constant along the Cu$'$ chains.} 
\label{tablecouplings}
\end{table}

The symmetry of the Hamiltonian (\ref{Hspeff}) allows for some
simplifications. First of all we use the equivalence of coupling to
the O2a and O2b displacements (see Fig.~\ref{unitcell}):
\begin{eqnarray}
g^{x}_{\mO}&=&g^{x}_{\mOb} = \ \ g^{x}_{\mOa},\\
g^{y}_{\mO}&=&g^{y}_{\mOb} = -g^{y}_{\mOa}.
\end{eqnarray}
From the symmetry of the T$_2^+$ modes (see Fig.~\ref{T2Modes}) we
see that the O2-$y$ components are in phase, 
i.e., $r^{y}_{{\mOa},{\bf n}}=r^{y}_{{\mOb},{\bf n}}=r^{y}_{{\mO},{\bf
n}}$, while the $x$ components exhibit an antiphase shift:
$-r^{x}_{{\mOa},{\bf n}}=r^{x}_{{\mOb},{\bf n}}=r^{x}_{{\mO},{\bf n}}$.

As indicated in Fig.~\ref{unitcell} we then cut the unit cell along
the $y$ axis in half separating the ions labeled ``prime'' from those
without a label. 
\begin{eqnarray}\label{rprimetrafo}
r^\alpha_{\nu',{\bf n}} &\rightarrow& -r^\alpha_{\nu,{\bf
n}+\hat{y}/2},
\\
{\bf S}'_{\bf n}\cdot{\bf S}'_{{\bf n}+\hat{z}} &\rightarrow&
{\bf S}_{{\bf n}+\hat{y}/2}\cdot{\bf S}_{{\bf n}+\hat{y}/2+\hat{z}}.
\label{Sprimetrafo}
\end{eqnarray}
The change of sign of the coordinates accounts for the antiphase
elongation of the two types of ions in the Peierls-active modes at the
wave vector of the instability ${\bf q}={\bf
q}_0=(\pi/a,0,\pi/c)$. Resummation ${\bf n}\to{\bf l}$ over all
the new cells, i.e., twice as many with a new cell length
$b/2$ in the $y$ direction, yields
\begin{eqnarray}\label{Hspeffkurz}
H^{\rm NN}_{sp}\!&=&\!\sum_{\bf l}\!\Bigg[
 g^z_{\mCu} \left(r^z_{\mCu,{\bf l}} - r^z_{\mCu,{\bf l}+\hat{z}}\right)
+ g^y_{\mGe} \left(r^y_{\mGe,{\bf l}-\hat{y}} + r^y_{{\mGe},{\bf
l}}\right)
                                                \nonumber\\&&\quad\ \
+\; g^x_{\mO} \left(r^x_{\mO,{\bf l}-\hat{y}}+r^x_{{\mO},{\bf l}}\right)
                                                \nonumber\\&&\quad\ \
+\; g^y_{\mO} \left(r^y_{\mO,{\bf l}-\hat{y}} + r^y_{{\mO},{\bf
l}}\right)
\Bigg]\ e^{i\pi l_y}\ {\bf S}_{\bf l}\cdot{\bf S}_{{\bf l}+\hat{z}}\,.
\end{eqnarray}
The overall change of sign in the first sum with respect to the second
in Eq.\ (\ref{Hspeff}) has been incorporated in the phase factor $e^{i\pi
l_y}$. This change of sign translates into the antiphase shift of the
spin-Peierls ordering between neighboring Cu chains in the
$y$-direction.

Now we substitute the displacements $r^\alpha_{\nu,{\bf l}}$ with the
${\bf q}$-space normal coordinates (\ref{xdecomp}). For clarity we
introduce the abbreviation
\begin{equation}\label{defSq}
S_{-{\bf q}} := \sum_{\bf l}\ e^{i{\bf q}{\bf R}_{\bf l}}\ e^{i\pi l_y}\
        {\bf S}_{\bf l}\cdot{\bf S}_{{\bf l}+\hat{z}}
\end{equation}
for the Fourier transform of the nearest-neighbor spin-spin
correlation operator:
\begin{equation}\label{HspQ}
H^{\rm NN}_{sp}=\frac{1}{\sqrt{N}}\sum_{\bf q}\ S_{-{\bf q}}\ \sum_\lambda\
                \sqrt{\frac{2\Omega_{\lambda,{\bf q}}}{\hbar}}\
                g_\lambda({\bf q})\ Q_{\lambda,{\bf q}}.
\end{equation}
Here the effective normal mode coupling constant
\begin{eqnarray}\label{geff}
\sqrt{\frac{2\Omega_{\lambda,{\bf q}}}{\hbar}}\ g_\lambda({\bf q}) 
&:=& (1 - e^{i q^z c})\ g_\mCu^z\ 
                \frac{{\rm e}_{\mCu}^z(\lambda,{\bf q})}{\sqrt{m_\mCu}}
\nonumber\\&&\hspace{-7ex}
+\, (e^{-i q^y b/2} + 1)
\Bigg( g_\mGe^y\
                \frac{{\rm e}_{\mGe}^y(\lambda,{\bf
q})}{\sqrt{m_\mGe}} 
\nonumber\\&&\hspace{-1ex}
+\, g_\mO^x\
                \frac{{\rm e}_{\mO}^x(\lambda,{\bf q})}{\sqrt{m_\mO}} +
        g_\mO^y\
                \frac{{\rm e}_{\mO}^y(\lambda,{\bf q})}{\sqrt{m_\mO}}
\Bigg)
\end{eqnarray}
was introduced.\cite{geology}

The next step is to transform the normal coordinates to boson creation
and annihilation operator representation via Eq.\ (\ref{Qtob}): 
\begin{equation}\label{Hspbose}
H^{\rm NN}_{sp}=\frac{1}{\sqrt{N}}\sum_{\bf q}\ S_{-{\bf q}}\ \sum_\lambda\
                g_\lambda({\bf q})\
\left(b^\dagger_{\lambda,-{\bf q}}+b^{}_{\lambda,{\bf q}}\right).
\end{equation}
This is the representation usually used in theoretical approaches to
spin-phonon coupling.\cite{Gros98CGO}
Since the polarization vectors are known for ${\bf q}_0$ (Table
\ref{tablefrequencies}), Eq.~(\ref{geff}) defines the relation between
the coupling to the linear atomic deviations $g_\nu^\alpha$ (Table
\ref{tablecouplings}) and the normal mode coupling constants
$g_\lambda({\bf q})$. 

The NNN exchange term $J_2\ {\bf S}_{\bf l}\cdot{\bf S}_{{\bf
l}+2\hat{z}}$ leads to a magnetoelastic coupling equivalent to the one
for the NN exchange shown in Eq.~(\ref{Hspeffkurz}).  Including all
ionic linear elongations contributing to the Cu-O2-O2-Cu NNN
superexchange path the prefactors in the resulting reciprocal-space
coupling constants --- compare Eq.\ (\ref{geff}) --- then are $(1 - e^{2
i q^z c})$ for the Cu part and $(1 + e^{i q^z c})$ for the other
ions. The coupling of the $J_2$ term vanishes at the wave vector of
the instability ${\bf q}={\bf q}_0 = (\pi/a,0,\pi/c)$ and does thus
not directly influence the transition. Within the present work we can
disregard this contribution.


\section{Coupling to bond angles and lengths}
\label{anglesection}

The two lower Peierls-active modes essentially
vary the angles  $\eta^\kappa=\angle$[Cu-O2$\kappa$-Cu] and
$\delta^\kappa=\angle$[O2$\kappa'$-O2$\kappa$-Ge]. 
Together with the bond lengths
$d^\kappa_\mCu=\overline{{\rm Cu-O2}\kappa}$ and
$d^\kappa_\mGe=\overline{{\rm Ge-O2}\kappa}$
they represent the natural set of coordinates of the lattice
vibrations in the irreducible group of the T$_2^+$ modes.
The index $\kappa\in\{a,a',b,b'\}$ was introduced to label the
position on the different oxygen atoms in the unit
cell.

Introducing the variable
$\theta^\kappa\in\{\eta^\kappa,\delta^\kappa,d^\kappa_\mCu,d^\kappa_\mGe\}$
we can write
\begin{equation}\label{DeltaJangle}
\Delta J_{{\bf l},{\bf l}+\hat{z}}=
\sum_{\{\theta\}}\frac{\partial J}{\partial
                        \theta^\kappa_{\bf n}} \Delta\theta^\kappa_{\bf n}=
\sum_{\{\theta\}} g_{\theta} \sum_{\nu}
        \frac{\partial \theta^\kappa_{\bf n}}{\partial
        r^\alpha_{\nu,{\bf n}}} r^\alpha_{\nu,{\bf n}}\,.
\end{equation}
Here we defined the coupling constants
$g_{\theta}=(\partial J)/(\partial \theta^\kappa_{\bf n})$, which are
independent of $\kappa$. For reasons of translational invariance we
can drop the unit-cell index ${\bf n}$. The linear coefficients of the
Taylor expansions $(\partial \theta^\kappa)/(r^\alpha_{\nu})$ at
different positions $\kappa$ in the unit cell all yield the same
numerical coefficients but with varying signs. The absolute values of
the coefficients are given in Table \ref{tabletaylor}.

\begin{table}[tb]
\renewcommand{\arraystretch}{1.3}
\begin{tabular}{cllll}
$\theta$ & 
\multicolumn{1}{c}{$\eta$} &
\multicolumn{1}{c}{$\delta$} &
\multicolumn{1}{c}{$ d_\mCu$} & 
\multicolumn{1}{c}{$ d_\mGe$} 
                                                          \\[0.2ex]\hline
$|\partial\theta^\kappa/\partial r^z_{\mCu}|$ &
   0.11 $\frac{\pi}{\mbox{\tiny\AA}}$ & 0 &
   0.76 & 0 \\
$|\partial\theta^\kappa/\partial r^y_{\mGe}|$ &  
   0 & 0.11 $\frac{\pi}{\mbox{\tiny\AA}}$ & 0
   & 0.82 \\
$|\partial\theta^\kappa/\partial r^x_{\mO}|$ & 
   0.21 $\frac{\pi}{\mbox{\tiny\AA}}$ & 
   0.29 $\frac{\pi}{\mbox{\tiny\AA}}$ & 0.54 & 0.57 \\
$|\partial\theta^\kappa/\partial r^y_{\mO}|$ & 
   0.14 $\frac{\pi}{\mbox{\tiny\AA}}$ & 
   0.32 $\frac{\pi}{\mbox{\tiny\AA}}$ & 0.36 & 0.82 \\
$\theta_0$ &
    0.55\,\ $\pi$ & 0.89\,\ $\pi$ & 1.93 \AA & 1.73 \AA
\end{tabular}
\caption{Linear coefficients of the expansion of the angles and bond
lengths as a function of the linear atomic elongations, as defined in
Eq.\ (\protect\ref{DeltaJangle}). The variables
$\kappa\in\{a,a',b,b'\}$ and 
$\theta^\kappa\in\{\eta^\kappa,\delta^\kappa,d^\kappa_\mCu,d^\kappa_\mGe\}$
are introduced in the text. The last
line holds the experimental equilibrium angles and bond
lengths (Ref.\ \protect\onlinecite{Braden96CGO})
[$\theta^\kappa=\theta_0+\Delta\theta^\kappa$, 
$\Delta\theta^\kappa$ is defined in Eq.~(\protect\ref{DeltaJangle})].}
\label{tabletaylor}
\end{table}

Considering all the relevant bonds and angles and using the
decomposition (\ref{DeltaJangle}) we can set up a
spin-phonon Hamiltonian similar to Eq.\ (\ref{Hspeff}) in the previous
section. By a simple comparison of the coefficients we obtain the
transformation matrix between the angular and bond-length coupling
constants and the linear atomic deviation coupling constants.
\begin{equation}\label{comparecoeff}
\left(\begin{array}{c}
g_\mCu^z \\ g_\mGe^y \\ g_\mO^x \\ g_\mO^y
\end{array}\right)\!\!=\!\!
\left(\begin{array}{r@{}lr@{}lr@{}lr@{}l}
 -&0.22\; \frac{\pi}{\mbox{\tiny\AA}} &  &0 & -&1.52 &  &0    \\
  &0  & -&0.11\; \frac{\pi}{\mbox{\tiny\AA}} &  &0    &  &0.82 \\
  &0.21\; \frac{\pi}{\mbox{\tiny\AA}} & 
  &0.29\; \frac{\pi}{\mbox{\tiny\AA}}       & -&1.08 &  &0.57 \\
 -&0.14\; \frac{\pi}{\mbox{\tiny\AA}} & 
  &0.32\; \frac{\pi}{\mbox{\tiny\AA}}       &  &0.72 & -&0.82
\end{array}\right)\!\!\!
\left(\begin{array}{c}
g_\eta \\ g_\delta \\ g^d_{\mCu} \\ g^d_{\mGe}
\end{array}\right)
\end{equation}
Together with Eq.\ (\ref{geff}) we now can
determine all coupling constants if any four of the them are known.


\section{Normal mode coupling constants}
\label{sectionnormal}

We now numerically determine the four normal mode
coupling constants. We have shown the RPA approach by Cross and Fisher
\cite{CF79} for the Hamiltonian
\begin{eqnarray}\label{CFH}
H&=&J\ \sum_{\bf l}\ {\bf S}_{\bf l}\cdot{\bf S}_{{\bf l}+\hat{z}}
+J_2\ \sum_{\bf l}\ {\bf S}_{\bf l}\cdot{\bf S}_{{\bf l}+2\hat{z}}
\nonumber\\&&
+\ \sum_{{\bf q},\lambda}\hbar\Omega_{\lambda,{\bf q}}
\left(b^\dagger_{\lambda,{\bf q}}b^{}_{\lambda,{\bf
q}}+\frac{1}{2}\right)
\nonumber\\&&
+\ \frac{1}{\sqrt{N}}\sum_{\bf q}\ S_{-{\bf q}}\ \sum_\lambda\
                g_\lambda({\bf q})\
\left(b^\dagger_{\lambda,-{\bf q}}+b^{}_{\lambda,{\bf q}}\right)
\end{eqnarray}
to satisfactorily describe the dynamics of the Peierls-active Phonon
modes.\cite{Gros98CGO} It consists of the Heisenberg chain
(\ref{Hs}), the harmonic phonon part (\ref{Hpbose}) and the
spin-phonon coupling term (\ref{Hspbose}) all discussed above.
We have given an expression for the critical temperature of the
spin-Peierls transition\cite{Gros98CGO} which we generalize to the
four Peierls-active modes we have to consider herein
($\Omega_{\lambda,{\bf q}_0}\equiv\Omega_\lambda$,
$g_\lambda({\bf q}_0)\equiv g_\lambda$).\cite{geology}
\begin{equation}
k_{\rm B} T_{\rm SP} = \left({2g_1^2\over\hbar\Omega_1}
              +{2g_2^2\over\hbar\Omega_2}
              +{2g_3^2\over\hbar\Omega_3}
              +{2g_4^2\over\hbar\Omega_4}\right)\chi_0
\label{T_SP}
\end{equation}
The factor $\chi_0\approx0.5$ is a contribution of the
static spin-polarization function at the appropriate wave vector.
Its value is controversial and we have adopted a mean of the proposed
values. Please refer to the discussion in Sec.\
\ref{sectiondiscussion} for the details.

\subsection{Ginzburg criterion}\label{sectionGinzburg}

The Ginzburg criterion gives an estimate of the temperature range of
the critical region in which fluctuations suppress the
applicability of mean-field approaches (or RPA). It is obtained
through comparing the theoretical correction of Gaussian
fluctuations to the specific heat
\begin{equation}\label{DeltaCp}
C_p - C_{p,0} = \frac{a\,b\,c}{16\pi}\,\frac{T_{\rm SP}^2}{(T-T_{\rm SP})^2}
\,\frac{k_{\rm B}}{\xi_a\xi_b\xi_c}
\end{equation}
with the experimental jump in the specific heat at the
transition.\cite{Landau} The correlation lengths $\xi_a$, $\xi_b$, and
$\xi_c$ along the respective crystallographic axes can be obtained
from fits to the diffuse x-ray data from Schoeffel {\it et
al.}\cite{Pouget96CGO} 
\begin{eqnarray*}
\xi_a&\approx&0.50\ a \ 
            \left[(T-T_{\rm SP})/T_{\rm SP}\right]^{-\frac{1}{3}},\\
\xi_b&\approx&0.65\ b \ 
            \left[(T-T_{\rm SP})/T_{\rm SP}\right]^{-\frac{1}{3}},\\
\xi_c&\approx&3.06\ c \ 
            \left[(T-T_{\rm SP})/T_{\rm SP}\right]^{-\frac{1}{3}}.
\end{eqnarray*}
The specific-heat jump has been determined by Lasjaunias {\it et
al.}\cite{Lasjaunias97CGO} to be $\Delta C_{\rm exp}=0.73 k_{\rm B}$
at $T_{\rm SP}$ per unit-cell volume. Requiring $C_p - C_{p,0} \ll \Delta
C_{\rm exp}$ we find the Ginzburg criterion to be 
\begin{equation}\label{Ginzburg}
(T-T_{\rm SP})\ \gg\ 0.03\, T_{\rm SP}\ =\ 0.4\ {\rm K}\,.
\end{equation}
In accordance with the mean-field approach to the susceptibility
by Kl\"umper {\it et al.}\cite{Kluemper98CGO}\ we conclude that beyond
a region of 3--4 K around  $T_{\rm SP}$ the mean-field theory is reliable.

\subsection{Mean-field approach}

The transition temperature given for CuGeO$_3$ with $14.1$ K, one
parameter is fixed through Eq.\ (\ref{T_SP}). As we shall discuss 
now, the others can be estimated from the 
polarization vectors of the Peierls-active phonon modes and the static
distortion in the dimerized phase at $4$ K also given by Braden {\it
et~al.}\ \cite{Braden96CGO} For the fixed wave vector of the Peierls
instability ${\bf q}_0$ we can derive from expressions (\ref{xdecomp})
and (\ref{Qtob}) a relation between a static lattice distortion
$\left\langle r^\alpha_\nu \right\rangle$ and the expectation values
of the displacement of the eigenmodes out of the harmonic
equilibrium $\left\langle b_{\lambda,{\bf q}_0} \right\rangle$:
\begin{equation}
\left\langle r^\alpha_{\nu} \right\rangle=
\frac{\left\langle r^\alpha_{{\bf q}_0,\nu} \right\rangle}{\sqrt{N}}=
        \sum_\lambda\
        \frac{{\rm e}^\alpha_\nu(\lambda,{\bf q}_0)}{\sqrt{Nm_\nu}}\
        \sqrt{\frac{2\hbar}{\Omega_{\lambda,{\bf q}_0}}}
\left\langle b_{\lambda,{\bf q}_0} \right\rangle .
\label{rtob}
\end{equation}
Introducing the canonical transformation
\begin{equation}\label{canonical}
\tilde{b}_{\lambda,{\bf q}}=b_{\lambda,{\bf q}}+\frac{1}{\sqrt{N}}
        \frac{g_\lambda({\bf q})}{\hbar\Omega_{\lambda,{\bf q}}}
        S_{\bf q}
\end{equation}
for the Bose annihilation and creation operators $b_{\lambda,{\bf
q}}$, where $S_{\bf q}$ was defined by Eq.\ (\ref{defSq}), the Hamiltonian
(\ref{CFH}) decouples into 
\begin{eqnarray}\label{CFHcanon}
H&=&J\ \sum_{\bf l}\ {\bf S}_{\bf l}\cdot{\bf S}_{{\bf l}+\hat{z}}
- \frac{1}{N}\sum_{\lambda\bf q}\ \frac{|g_\lambda({\bf q})|^2}
         {\hbar\Omega_{\lambda,{\bf q}}}
         S_{-{\bf q}}S_{{\bf q}}
\nonumber\\&&\qquad\qquad
+\; \sum_{{\bf q},\lambda}\hbar\Omega_{\lambda,{\bf q}}
\left(\tilde{b}^\dagger_{\lambda,{\bf q}}
      \tilde{b}^{}_{\lambda,{\bf q}}+\frac{1}{2}\right)\,.
\end{eqnarray}
The operators $\tilde{b}_{\lambda,{\bf q}}$ do not satisfy Bose
commutation relations and
since $[S_{-{\bf q}},\tilde{b}^{}_{\lambda,{\bf q}}]_-\neq 0$ the
solution of this Hamiltonian is not at all evident. But in a
mean-field-like approach we can assume
$\langle \tilde{b}^{}_{\lambda,{\bf q}}\rangle=0$ so that
from Eq.\ (\ref{canonical}) follows
\begin{equation}\label{expectb}
\left\langle b_{\lambda,{\bf q}} \right\rangle=
-\frac{1}{\sqrt{N}}
        \frac{g_\lambda({\bf q})}{\hbar\Omega_{\lambda,{\bf q}}}
        \left\langle S_{\bf q} \right\rangle .
\end{equation}
The mean-field ansatz is reasonable here, since we are interested in
temperatures of 4 K which is far from the critical region and the
dimerization is as good as saturated.\cite{Hirota95CGO}

\subsection{Values}

With Eqs.\ (\ref{rtob}) and (\ref{expectb}) we are left with a set of
linear equations. The values of the frequencies $\Omega_\lambda$ are
given in Table \ref{tablefrequencies}, the polarization vectors
${\rm e}^\alpha_\nu(\lambda,{\bf q}_0)$ enter via the matrix {\bf M}
defined in Eq.\ (\ref{eMatrix}), and $\langle {\bf r} \rangle_{T=4{\rm
K}}$ is given in Eq.~(\ref{rVector}).
\begin{equation}\label{LGS1}
\langle {\bf r} \rangle_{T=4{\rm K}}
=-\frac{\langle S_{{\bf q}_0}\rangle}{N}\sqrt\frac{2}{\hbar}
\ {\bf M}\left(\begin{array}{c}
g_1/\sqrt{\Omega^3_1}  \\
g_2/\sqrt{\Omega^3_2}  \\
g_3/\sqrt{\Omega^3_3}  \\
g_4/\sqrt{\Omega^3_4}
\end{array}\right)
\end{equation}
The solution of the equations gives the coupling constants as a
function of $N/\langle S_{{\bf q}_0}\rangle$. The latter is
then determined by the critical temperature $T_{\rm SP}=14.1$ K via
Eq.~(\ref{T_SP}).
\begin{equation}\label{S/N}
\frac{\langle S_{{\bf q}_0}\rangle}{N}=\frac{1}{N}\sum_{\bf l}
(-1)^{l_x+l_y+l_z}
\langle{\bf S}_{\bf l}\cdot{\bf S}_{{\bf l}+\hat{z}}\rangle
=0.59
\end{equation}
For a spin-$1/2$ system with two Cu chains per unit cell we have
$\langle S_{{\bf q}_0}\rangle/N \le 0.75$ where $0.75$ is reached in
the fully dimerized state. In the uniform Heisenberg case $\langle
S_{{\bf q}_0}\rangle/N=0$.

In Table \ref{tablenormalmode} we show the calculated coupling
constants of the spin system to the Peierls-active eigenmodes of the
lattice at the wave vector of the instability ${\bf q}_0$. The signs
are such that all contributions in the spin-phonon coupling term in
the Hamiltonian (\ref{CFH}) are negative when the phonon modes are
macroscopically occupied as determined via Eq.~(\ref{rtob}) in Sec.\ 
\ref{sectiondimer}. The mode at $\Omega_2/(2\pi)=6.5$ THz is dominant,
by its symmetry it essentially varies the angles $\eta$. This will be
reflected in the corresponding coupling constant discussed below. 

Note that the influence on the transition temperature of the lowest
$\Omega_1/(2\pi)=3.1$ THz mode is as important as that of
$\Omega_3/(2\pi)=11$ THz, because of the frequencies in the
denominator of Eq.\ (\ref{T_SP}).

\begin{table}[tb]
\begin{tabular}{cccc}
$g_1/k_{\rm B}$ & $g_2/k_{\rm B}$ & $g_3/k_{\rm B}$ & $g_4/k_{\rm B}$
\\\hline
-15 K & 58 K & -30 K & -12 K
\end{tabular}
\caption{Normal mode coupling constants for the four Peierls-active
phonon modes at ${\bf q}_0$, determined by Eq.\ (\protect\ref{LGS1}).}
\label{tablenormalmode}
\end{table}


\section{Microscopic coupling constants}

The numerical values of the normal mode coupling constants thus given,
the microscopic coupling constants can be determined. Using the matrix
(\ref{eMatrix}) we rewrite expression (\ref{geff}) for ${\bf q}={\bf
q}_0$ as 
\begin{equation}\label{LGS2}
{\bf M}^{\rm T}\
\left(\begin{array}{c}
g_\mCu^z \\ g_\mGe^y \\ g_\mO^x \\ g_\mO^y
\end{array}\right)
=
\frac{1}{\sqrt{2\hbar}}\left(\begin{array}{c}
g_1\,\sqrt{\Omega_{1}} \\
g_2\,\sqrt{\Omega_{2}} \\
g_3\,\sqrt{\Omega_{3}} \\
g_4\,\sqrt{\Omega_{4}} \\
\end{array}\right)
\,,
\end{equation}
and compute the coupling to the linear atomic elongations. Then we
calculate the angular and bond length couplings using 
Eq.~(\ref{comparecoeff}). The resulting values are given in Tables
\ref{tablelinear} and \ref{tableangular}, respectively.

\begin{table}[tb]
\begin{tabular}{cccc}
$g_\mCu^z/k_{\rm B}$ & $g_\mGe^y/k_{\rm B}$ &
$g_\mO^x/k_{\rm B}$ & $g_\mO^y/k_{\rm B}$
\\\hline
-890 K/\AA & -110 K/\AA & 400 K/\AA  & -91 K/\AA
\end{tabular}
\caption{Coupling constants for the linear atomic displacements
calculated via Eq.\ (\protect\ref{LGS2}) using the values for
$g_\lambda$ from Table \protect\ref{tablenormalmode}.}
\label{tablelinear}
\end{table}

\begin{table}[tb]
\begin{tabular}{cccc}
$g_\eta/k_{\rm B}$ & $g_\delta/k_{\rm B}$ &
$g^d_{\mCu}/k_{\rm B}$ & $g^d_{\mGe}/k_{\rm B}$
\\\hline
15 K/deg & 1.5 K/deg & 180 K/\AA & -96 K/\AA
\end{tabular}
\caption{Coupling constants for the angles and bond lengths calculated
via Eq.\ (\protect\ref{comparecoeff}) using the values for
$g^\alpha_\nu$ from Table \protect\ref{tablelinear}.}
\label{tableangular}
\end{table}

The results allow
for some immediate conclusions:
\\
$\bullet$ The coupling to $\eta$, i.e., $g_\eta$, is the dominant
contribution. 
\\
$\bullet$ The signs of the coupling constants are correct, $J$
increases with increasing angles and decreasing O2-Ge bond length. The
positive value of $g^d_{\mCu}$ indicates that the
ferromagnetic exchange is weakened more than the antiferromagnetic
exchange when stretching the O2-Cu bond. This is consistent with the net
ferromagnetic exchange of the O2-Cu plaquettes without the germanium
side group predicted by Geertsma and Khomskii.\cite{Geertsma96CGO}
\\
$\bullet$ Variation of the coupling constants shows $g_1$ to
couple mainly to the angles $\delta$,
$g_2$ to $\eta$, and $g_3$ and
$g_4$ to be almost entirely bond stretching.\cite{Braden98CGO}
While the results for $g_\eta$ and $g_\delta$ are robust under
variation of the parameters, the values of $g^d_\mCu$ and $g^d_\mGe$
are less fixed within the accuracy of our approach.
\\
$\bullet$
From magnetostriction data B\"uchner {\it
et~al.}\cite{Buechner98CGO}\ expect the influence of the Cu-O2-Cu angle
$\eta$ on the magnetic exchange to be of the order of $2\,\partial
J/(J\partial \eta^\kappa) \approx 10$\% per degree, and for the
O2-O2-Ge angle their value is $2\, \partial  J/(J\partial
\delta^\kappa) \approx 1$\%. For $J/k_{\rm B}=150$ K we obtain about
twice the values (see Table \ref{summaryJofangle}).
\\
$\bullet$ Comparing $g^d_{\mGe}$ and $g_\mGe^y$ shows the
effect of the germanium elongation on the magnetic exchange to be due
mainly to the stretching of the O2-Ge bond. The contribution of the 
Ge side group to the magnetic exchange should depend on the O2-O2-Ge
angle as $J_{\mbox{\tiny side}}\sim\cos\delta$.
Therefore, the angle $\delta^\kappa\approx 160^\circ=0.89\pi$ being
close to $\pi$, the angular dependency of $J$ on
$\delta$ is quite small in spite of the large entire side-group
effect, which is of similar magnitude as that of the CuO$_4$ plaquette
elongation.\cite{Geertsma96CGO,Braden96CGO}
\\
$\bullet$
Two groups analyzed the structural dependence of the superexchange
within similar microscopic models. Geertsma and Khomskii
\cite{Geertsma96CGO} obtained $J/k_{\rm B}=135$ K and found 
$2\,\partial J_{\rm\tiny geo}/(J\partial \eta^\kappa) \approx 16$\% and 
$2\,\partial J_{\rm\tiny geo}/(J\partial \delta^\kappa)
\approx 0.6$\% per degree. These values only accout for the
``geometrical'' contribution and are thus lower bounds. Braden {\it et
al.}\cite{Braden96CGO}\ found $J/k_{\rm B}=160$ K and gave $2\,\partial
J/(J\partial \eta^\kappa) \approx 44$\% and $2\,\partial J/(J\partial
\delta^\kappa) \approx 1.1$\% per degree. The agreement between the
microscopic models is affected by the choice of the parameters and the
number of orbitals taken into consideration. 

A summary of the values obtained in the different approaches is given
in Table \ref{summaryJofangle}.
\begin{table}[bt]
\begin{tabular}{lrr}
\multicolumn{1}{c}{Method [Reference]} &
\multicolumn{1}{c}{$\frac{\partial J}{J\partial \eta^\kappa}$} &
\multicolumn{1}{c}{$\frac{\partial J}{J\partial \delta^\kappa}$}
\\\hline
Harm.\ theory and mean-field [here] $\!\!\!\!$ & 10\%$\frac{1}{\rm deg}\ \ \ $
& 1\%$\frac{1}{\rm deg}\ \ \ $ \\
Microscopic superexchange [\hspace*{-0.8ex}\onlinecite{Geertsma96CGO}] &
            $\ge 8$\%$\frac{1}{\rm deg}\ \ \ $ &
            $\ge 0.3$\%$\frac{1}{\rm deg}\ \ \ $ \\
Microscopic superexchange [\hspace*{-0.8ex}\onlinecite{Braden96CGO}] &
            22\%$\frac{1}{\rm deg}\ \ \ $ &
            0.6\%$\frac{1}{\rm deg}\ \ \ $ \\
Pressure vs.\ magnetostriction
[\hspace*{-0.8ex}\onlinecite{Buechner98CGO}]
& $\sim 5$\%$\frac{1}{\rm deg}\ \ \ $
& $\sim 0.5$\%$\frac{1}{\rm deg}\ \ \ $
\end{tabular}
\caption{Variation of $J$ with the variation of the angles. Note that
in our notation there are two angles $\eta^\kappa$ and two angles
$\delta^\kappa$ contributing each to the Cu-Cu superexchange path (see
Fig.\ \protect\ref{unitcell}).} 
\label{summaryJofangle}
\end{table}


\section{Static distortion}
\label{sectionstatic}

The microscopic coupling constants given, we can directly calculate the effect
of static distortions of the lattice geometry on the magnetic exchange.

\subsection{Dimerization}\label{sectiondimer}

Using the static displacements of the ions in the spin-Peierls phase at
$T=4$ K,\cite{Braden96CGO} one may calculate the alternation of the
magnetic exchange usually used in mean-field approaches to the
spin-phonon coupling, i.e., 
\begin{equation}\label{Hdelta}
H_{\mbox{\tiny MF}}=J\ \sum_{l_z}\ \left(1 + (-1)^{l_z}\delta_J\right)\
        {\bf S}_{l_z}\cdot{\bf S}_{l_z+1}\,.
\end{equation}
This is achieved by substituting in the spin-phonon coupling term
(\ref{Hspeffkurz}) the atomic displacements by their expectation values
$r_{\nu,{\bf l}}^\alpha \to (-1)^{l_z+l_x}\langle r_\nu^\alpha
\rangle_{T=4{\rm K}}$ and comparing the resulting
$\langle H^{\rm NN}_{sp} \rangle_{T=4{\rm K}}$ with Eq.\ (\ref{Hdelta}).
Equivalently one can calculate $\langle H^{\rm NN}_{sp} \rangle_{T=4{\rm K}}$
by using the static angular and bond length
deviations\cite{Braden96CGO} yielding the same results.

We find $\delta_J J/k_{\rm B} = 17$ K or $\delta_J\approx0.11$
($J/k_{\rm B}=150$ K). By solving the system of linear equations
defined by Eq.\ (\ref{rtob}) for ${\bf q}={\bf q}_0$, the expectation
values $\langle b_\lambda \rangle/\sqrt{N}$ have been determined to be
$0.061$, $-0.11$, $0.034$, and $0.006$ for $\lambda=1$, $2$, $3$, and
$4$, respectively. The elastic energy per unit cell of the spin-Peierls
distortion at $T\sim4$ K then is given by  
\begin{equation}\label{E_p}
{\langle H_p \rangle \over N\,k_{\rm B}}=
            \sum_\lambda\frac{\hbar\Omega_\lambda}{N\,k_{\rm B}}\
            \langle b_\lambda \rangle^2
       = 5\ {\rm K}\,.
\end{equation}
This energy loss has to be compensated by the spin system.
Considering that the maximum gain of magnetic energy is reached in the
fully dimerized case with $0.375\, \delta_J J$ per Cu ion, we find a
lower boundary for the dimerization of $\delta_J > 0.044$.
Including a NNN term in Eq.\ (\ref{Hdelta}) with $J_2/J=0.241$ as
studied by Chitra {\it et al.}\cite{Chitra95DMRG} using a density
matrix renormalization group (DMRG) approach we find $\delta_J \ge
0.078$.

Our result is within a factor of two of the values obtained by
using $\partial J / \partial \eta$ and $\partial J / \partial
\delta$ obtained from the magnetostriction results\cite{Buechner98CGO}
and from the microscopic models.\cite{Geertsma96CGO,Braden96CGO} 
All other published estimates of the dimerization result from an
analysis of the magnetic excitation spectra observed by inelastic
neutron or Raman scattering. Most of these estimates are based on
the static dimerized Hamiltonian
(\ref{Hdelta}) (Refs.\ \onlinecite{Riera95CGO,Castilla95CGO}) 
and yield dimerization values much smaller than the one reported here
(see Table \ref{summarydelta}). Augier and Poilblanc\cite{Augier98CGO}
as well as Wellein {\it et al.}\cite{Wellein98CGO}\ extend the static
model by coupling to dynamical phonons which reduces the magnetic gap
by lowering the effective lattice distortion acting on 
the spin system.\cite{Buechner98CGO} The derivation of their model and
the significance of the phonon dynamics are more closely discussed in
Sec.\ \ref{sectionrealspace}. Introducing interchain coupling may 
further suppress the spin gap.\cite{Uhrig97CGO} For an extensive
discussion, see Ref.\ \onlinecite{Brenig97CGO}. 

All methods incorporate more or less crude approximations to the real
physical situation leaving the question of the true value of $\delta_J$
unanswered. Our lower boundary should be rather reliable though. The
values obtained in the different approaches are given 
in Table \ref{summarydelta} for comparison. 
\begin{table}[bt]
\begin{tabular}{lrr}
\multicolumn{1}{c}{Method [Reference]} &
\multicolumn{1}{c}{$\delta_J$} \\\hline
Harmonic theory and mean-field [here] & 0.11 \\
Macroscopic occupation of T$_2^+$ modes [here] & $>$0.04 \\
Microscopic superexchange [\hspace*{-0.8ex}
        \onlinecite{Geertsma96CGO,Braden96CGO}] &  0.07 to 0.2\\
Dynamic phonons and experimental gap
                  [\hspace*{-0.8ex}\onlinecite{Buechner98CGO}]
                             & $\sim 0.05$ \\
Static phonons and experimental gap
[\hspace*{-0.8ex}\onlinecite{Riera95CGO,Castilla95CGO}]
                             & 0.01 to 0.03 \\
Coupled chains [\hspace*{-0.8ex}\onlinecite{Brenig97CGO}]
                             & 0.01 to 0.12 \\
\end{tabular}
\caption{Exchange-alternation in $J[1+(-1)^{l_z}\delta_J]$.}
\label{summarydelta}
\end{table}

\subsection{Pressure}

Br\"auninger {\it et al.}\ \cite{Braeuninger97CGO} and
Braden {\it et al.}\ \cite{Bradentbp} have investigated
the pressure dependence of the angles and bond lengths in CuGeO$_3$
under hydrostatic pressure. The linearity of the pressure dependence
is reasonable for pressures $<2$ GPa. The values for the pressure
gradients obtained from Ref.\ \onlinecite{Bradentbp} are shown in
Table \ref{tablePgradient}. 

Regarding the partial derivative of the exchange integral
$\partial J_\theta/\partial p=(\partial J/\partial \theta)
(\partial \theta/\partial p)$ 
we find immediately the pressure gradients of the different angular
and bond length contributions to $J$ as given in Table
\ref{tablePgradient}.
\begin{table}[tb]
\renewcommand{\arraystretch}{1.4}
\begin{tabular}{ccccc}
$\theta$ & $\eta$ & $\delta$ & $d_{\mCu}$ & $d_{\mGe} $ \\\hline
$\frac{\partial \theta^\kappa}{\partial p}$ &
$-0.16\ \frac{\rm deg}{\rm GPa}$ & 
$-1.3\ \frac{\rm deg}{\rm GPa}$ &
$-0.0012\ \frac{\mbox{\tiny \AA}}{\rm GPa}$ &
$-0.0033\ \frac{\mbox{\tiny \AA}}{\rm GPa}$ \\[1ex]
$\frac{\partial J_\theta}{\partial p}$ &
$-2.5\ \frac{k_{\rm B}{\rm K}}{\rm GPa}$ &
$-1.9\ \frac{k_{\rm B}{\rm K}}{\rm GPa}$ &
$-0.22\ \frac{k_{\rm B}{\rm K}}{\rm GPa}$ &
$0.32\ \frac{k_{\rm B}{\rm K}}{\rm GPa}$
\end{tabular}
\caption{Linear pressure gradients of angles and bond lengths from
experimental data in Ref.\ \protect\onlinecite{Bradentbp} (top) and
the resulting theoretical pressure gradients 
$\partial J_\theta/\partial p=(\partial J/\partial \theta)
(\partial \theta/\partial p)$ (bottom). The values for $\partial
J/\partial \theta$ are given in Table \protect\ref{tableangular}.} 
\label{tablePgradient}
\end{table}

Considering all four contributions, we obtain the total variation of
the antiferromagnetic exchange.
\begin{equation}\label{totalJ}
\frac{\partial J}{\partial p}=
2\frac{\partial J_\eta}{\partial p}+
2\frac{\partial J_\delta}{\partial p}+
2\frac{\partial J_\mGe}{\partial p}+
4\frac{\partial J_\mCu}{\partial p}=-9\ \frac{k_{\rm B}{\rm K}}{\rm
GPa}
\end{equation}

For $J/k_{\rm B}=150$ K this value corresponds to 
$\partial  J/(J\partial p) \approx -6$\% per GPa. The pressure
dependency of the magnetic susceptibility is directly related to the
magnetostriction. A value of $-\partial  \chi/(\chi\partial p) \sim
\partial  J/(J\partial p) \approx -8$\% per GPa was obtained after
averaging the uni-axial components.\cite{Lorenz96CGO,Fabricius98CGO}
Takahashi {\it et al.}\ \cite{Takahashi95CGO} have measured the
pressure dependence of the Curie constant $C$ by fitting a
Curie-Weiss law to the high-temperature tail of the magnetic
susceptibility. Assuming $C\sim 1/J$ one can estimate a value of about
$\partial J/(J\partial p) \approx -7$\% per GPa. Nishi and co-workers 
\cite{Nishi95CGO} compared fits to the dispersion of the lowest
triplet excitations at different pressures. They assume the ratio
between the exchange $J$ and next-nearest-neighbor exchange $J_2$ with
a value of $J_2/J\approx0.25$ not to alter under pressure, and find
$\partial J/(J\partial p)\approx -10$\% per GPa. In contrast to that
Fabricius {\it et al.}\ \cite{Fabricius98CGO} find  $J_2$ not to alter
under pressure. Then the  result from Nishi's analysis is corrected to
$\partial J/(J\partial p)\approx -8$\% per GPa. A summary of the
values is given in Table \ref{summaryJofP} showing their consistency.
\begin{table}[bt]
\begin{tabular}{lrr}
\multicolumn{1}{c}{Method [Reference]} &
\multicolumn{1}{c}{$\frac{\partial J}{J\partial p}$} \\\hline
Harmonic theory and mean-field [here] &
-6\% $\frac{1}{\rm GPa}$ \\
Susceptibility via magnetostriction
                      [\hspace*{-0.8ex}\onlinecite{Lorenz96CGO}] &
-8\% $\frac{1}{\rm GPa}$ \\
Curie-Weiss fit to the susceptibility
                      [\hspace*{-0.8ex}\onlinecite{Takahashi95CGO}] &
-7\% $\frac{1}{\rm GPa}$ \\
Fit to the triplet dispersion [\hspace*{-0.8ex}\onlinecite{Nishi95CGO}]
&
-8 to -10\% $\frac{1}{\rm GPa}$
\end{tabular}
\caption{Variation of $J$ with pressure.}
\label{summaryJofP}
\end{table}

\subsection{Thermal expansion and spontaneous strain}
\label{sectionthermal}

In a harmonic lattice the coefficients of linear thermal expansion
$\alpha=(\partial L)/(L\partial T)_p$ vanish. Here $L$ is the length
of the crystal in a given spatial direction.
Anharmonic contributions result in temperature-dependent phonon
frequencies which in turn yield finite values for $\alpha$. The
coefficient of thermal expansion is linked to the specific heat via the
(temperature dependent) Gr\"uneisen parameter. This implies in the
limiting cases $T\to 0:\ \alpha \sim T^3$ and $T\gg\Theta_D:\
\alpha\sim\mbox{constant}$, where $\Theta_D$ is the Debye
temperature.\cite{Ashcroft} 

The thermal expansion in CuGeO$_3$ can be attributed to two effects:
the usual anharmonic behavior described above and anomalies due to the
spin-phonon coupling.\cite{Braden98CGOanh,Winkelmann95CGO}
The coefficient of linear thermal expansion of the $c$ axis in CuGeO$_3$
has a negative sign between $T_{\rm SP}$ and $T\sim200$ K. The expansion 
of the $c$ axis enlarges $J$ via the angle $\eta$. The spin
system then gains energy when the temperature is lowered to $T \sim J$
by driving the anomaly.\cite{Sandvik98QMC} A rough quantitative
estimation can be extracted from the analysis of the temperature
dependence of the herein considered bond lengths and 
angles given by Braden {\it et al.}\cite{Braden98CGOanh} Their
temperature dependence between 295 and 20 K is close to linear and 
presented in Table \ref{tableTgradient}. Summing up the different
contributions equivalently to Eq.\ (\ref{totalJ}) yields $\partial
J/(J\partial T)\approx -2.6$\% per 250 K ($J/k_{\rm B}=150$ K). This
effect is a superposition of the normal thermal expansion with
positive $\partial J_{\rm norm}/(J\partial T)$ and the anomalous
effect at low temperature which can be estimated by $\partial J_{\rm
an}/(J\partial T) \le 2\,\partial J_\eta/(J\partial T)=-4.1$\% per 250
K.

\begin{table}[tb]
\renewcommand{\arraystretch}{1.4}
\begin{tabular}{ccccc}
$\theta$ & $\eta$ & $\delta$ & $d_{\mCu}$ & $d_{\mGe} $ \\\hline
$\frac{\partial \theta^\kappa}{\partial T}$ &
$-0.2\ \frac{\rm deg}{\rm 250 K}$ & 
$0.6\ \frac{\rm deg}{\rm 250 K}$ &
$0.0002\ \frac{\mbox{\tiny \AA}}{\rm 250 K}$ &
$-0.002\ \frac{\mbox{\tiny \AA}}{\rm 250 K}$ \\[1ex]
$\frac{\partial J_\theta}{\partial T}$ &
$-3.1\ \frac{k_{\rm B}{\rm K}}{\rm 250 K}$ &
$0.9\ \frac{k_{\rm B}{\rm K}}{\rm 250 K}$ &
$0.04\ \frac{k_{\rm B}{\rm K}}{\rm 250 K}$ &
$0.2\ \frac{k_{\rm B}{\rm K}}{\rm 250 K}$
\end{tabular}
\caption{Experimental linear temperature gradients of angles and bond
lengths from Ref.\ \protect\onlinecite{Braden98CGOanh} (top) and the
resulting theoretical contributions to the temperature dependence of
$J$ (bottom) between 20 and 295 K.}
\label{tableTgradient}
\end{table}

As the crystal undergoes the SP transition spontaneous strain appears
along all three orthorhombic
directions.\cite{Lorenz97CGO,Winkelmann95CGO} The strain couples  
different T$_2^+$ modes\cite{BruceCowley81} and gives a 
correction to Eq.~(\ref{LGS1}) which we now show to be
unimportant. The elastic energy per unit cell related to the
spontaneous strain at $T\sim4$ K can be estimated from the elastic
constants. The diagonal elastic constants were taken from the
ultrasound study by Poirier 
{\it et al.},\cite{Poirier95CGO} and off-diagonal terms were
calculated with the lattice-dynamical model\cite{Bradentbp} as shown
in Table \ref{tableelastic} using standard notation.\cite{Ashcroft}
With the values for the strain $\epsilon_i$ given by Winkelmann {\it
et al.}\cite{Winkelmann95CGO} we find
\begin{equation}
{E_{\rm strain} \over k_{\rm B}}=\frac{a\cdot b\cdot c}{2k_{\rm B}}
            \sum_{i,j=1,2,3} \epsilon_i C_{ij} \epsilon_j 
            = 7\cdot 10^{-4}\ {\rm K}\,.
\end{equation}
Note that the strain components $\epsilon_4=\epsilon_5=\epsilon_6$ vanish,
since the orthorhombicity is conserved. The elastic energy involved in
the strain is four orders of magnitude smaller than the elastic energy
of the dimerization given in Eq.~(\ref{E_p}).  

Note that the components of the spontaneous
strain\cite{Winkelmann95CGO} have the opposite sign compared with the
anomalies of the thermal expansion\cite{Braden98CGO} discussed
above. The spontaneous strain may thus be interpreted as a relaxation
of the latter when the spin system changes its character at the
spin-Peierls transition. The relaxation is of the order of 1\%.

\begin{table}[tb]
\renewcommand{\arraystretch}{1.4}
\begin{tabular}{lccccccccc}
   & $C_{11}$ & $C_{22}$ & $C_{33}$ & $C_{12}$ & $C_{13}$ & $C_{23}$ 
   \\\hline
Expt.\ $\left[10^{11}\frac{\rm dyn}{{\rm cm}^2}\right]$
& 7.4 & 2.1 & 33.2 &  &  &   \\
Theory\ $\left[10^{11}\frac{\rm dyn}{{\rm cm}^2}\right]$
& 8.2 & 5.0 & 34.6 & 3.0 & 4.0 & 2.2 
\end{tabular}
\caption{Experimental uniaxial elastic constants from Ref.\
\protect\onlinecite{Poirier95CGO} (top) and the theoretical
elastic constants obtained from the shell model (bottom).}
\label{tableelastic}
\end{table}


\section{Coupling constants for real-space normal coordinates}
\label{sectionrealspace}

In order to obtain real-space expressions we use the Fourier
representation of the Bose operators 
\begin{equation}\label{bltobq}
b_{\lambda,{\bf q}}=\frac{1}{\sqrt{N}}\sum_{\bf l}
        e^{-i{\bf q}{\bf R}_{\bf l}}\ b_{\lambda,{\bf l}}\,.
\end{equation}
For simplicity we neglect the wave-vector dependence of the
frequencies $\Omega_\lambda = \Omega_{\lambda,{\bf q}_0}$ and of the
polarization vectors ${\rm e}_{\nu}^\alpha(\lambda)
={\rm e}_{\nu}^\alpha(\lambda,{\bf q}_0)$. The coupling constants
$g_\lambda({\bf q}_0)$ in Eq.\ (\ref{geff}) then are divided into
\begin{eqnarray}
\sqrt{\frac{2\Omega_{\lambda}}{\hbar}}
g^\mCu_\lambda&=&g_\mCu^z
        \frac{{\rm e}_{\mCu}^z(\lambda)}{\sqrt{m_\mCu}}\,,
\label{gCu}
\\
\sqrt{\frac{2\Omega_{\lambda}}{\hbar}}
g^\loc_\lambda&=&g_\mGe^y
                \frac{{\rm e}_{\mGe}^y(\lambda)}{\sqrt{m_\mGe}} +
        g_\mO^x
                \frac{{\rm e}_{\mO}^x(\lambda)}{\sqrt{m_\mO}} +
        g_\mO^y
                \frac{{\rm e}_{\mO}^y(\lambda)}{\sqrt{m_\mO}}\,.
\label{gloc}
\end{eqnarray}
Transforming the Hamiltonian (\ref{CFH}) via Eq.\ (\ref{bltobq}) we
obtain in real space
\begin{eqnarray}\label{Hrbose}
H_{\mbox{\small r}}&=&\sum_{\lambda,{\bf
l}}\,\hbar\Omega_\lambda\,\left(
        b^\dagger_{\lambda,{\bf l}}b^{}_{\lambda,{\bf l}}
        +\frac{1}{2}\right)
+J\sum_{\bf l}\, {\bf S}_{\bf l}\cdot{\bf S}_{{\bf l}+1}
\nonumber\\&&
+\sum_{\lambda,{\bf l}}\;(-1)^{l_y}\;\Big[
   g^\mCu_\lambda
        \left(b^\dagger_{\lambda,{\bf l}}+b^{}_{\lambda,{\bf l}}
        -b^\dagger_{\lambda,{\bf l}+\hat{z}}
        -b^{}_{\lambda,{\bf l}+\hat{z}}\right)
\nonumber\\&&
  +\;g^\loc_\lambda
\left(b^\dagger_{\lambda,{\bf l}}+b^{}_{\lambda,{\bf l}}+
b^\dagger_{\lambda,{\bf l}+\hat{y}}+b^{}_{\lambda,{\bf
l}+\hat{y}}\right)
\Big]\,{\bf S}_{\bf l}\cdot{\bf S}_{{\bf l}+\hat{z}}\,.
\nonumber\\
\end{eqnarray}
The coupling constants are given in Table \ref{tablereal}. This result
implies that the oxygen and germanium displacements are of the same
importance for the spin-phonon coupling as the copper elongation. 
\begin{table}[tb]
\begin{tabular}{lrrrr}
$\lambda$ & \multicolumn{1}{c}{1} & \multicolumn{1}{c}{2} &
        \multicolumn{1}{c}{3} & \multicolumn{1}{c}{4} \\\hline
$g^\mCu_\lambda/k_{\rm B}$ & -0.5 K & 17 K & -20 K & 1.6 K \\
$g^\loc_\lambda/k_{\rm B}$ & -7.2 K & 12 K & 4.8 K & -7.5 K
\end{tabular}
\caption{Coupling constants to real-space normal modes obtained from
Eqs.\ (\ref{gCu}) and (\ref{gloc}).}
\label{tablereal}
\end{table}

Motivated by the symmetry of the Peierls-active phonon modes an
effective one-di\-men\-sional model can be obtained by restricting the
sum to a single chain. The Fourier transform of the one-dimensional
model derived from Eq.~(\ref{Hrbose}) shows the different 
$q$ dependences ($q\equiv q^z$) of the copper and the local term.
\begin{eqnarray}\label{Hr1DFourier}
H_{\mbox{\tiny 1D}}&=&\sum_{\lambda,{q}}\,\hbar\Omega_\lambda\,\left(
        b^\dagger_{\lambda,{q}}b^{}_{\lambda,{-q}}
        +\frac{1}{2}\right)
+J\sum_{l_z}\, {\bf S}_{l_z}\cdot{\bf S}_{{l_z}+1}
\nonumber\\&&
+\sum_{\lambda,{q}}
\frac{g_{\lambda\mbox{\tiny 1D}}(q)}{\sqrt{N}}
\left(b^\dagger_{\lambda,{-q}}+b^{}_{\lambda,{q}}\right)
\sum_{l_z}e^{iqR_{l_z}}\
              {\bf S}_{l_z}\cdot{\bf S}_{{l_z}+1}
\nonumber\\
\end{eqnarray}
Here we defined the 1D coupling constant
\begin{equation}\label{qprefactor}
g_{\lambda\mbox{\tiny 1D}}(q)=
            (1-e^{iqc}) g^\mCu_\lambda + 2g^\loc_\lambda\,.
\end{equation}

Several studies\cite{Augier98CGO,Wellein98CGO,Uhrig98SP,Sandvik98QMC}
have been carried out using real-space Hamiltonians in the form of
Eq.\ (\ref{Hrbose}) reduced to a one-dimensional model. 
Usually a single-mode Hamiltonian, only keeping the local
term is considered, i.e., in their notation $2g^\loc_1\equiv g$, while
the other coupling constants are set to zero. Considering $g^\mCu_2$, 
$g^\loc_2$, and $g^\loc_1$ being of the same order of magnitude, this
simplification should only yield qualitative results.

Yet, these treatments include the dynamics of the phonons. The
significance of the latter can be estimated from the size of the
zero-point motion of the ions. Without the negligible contribution
from the macroscopic occupation (Sec.\ \ref{sectiondimer}) the
fluctuations of the T$^+_2$ modes at $T=0$ can be obtained from
Eq.~(\ref{xdecomp}) using the approximation of dispersionless phonons
introduced above: 
\begin{equation}
\langle \overline{ (r^\alpha_\nu)^2} \rangle=
        \frac{1}{N}\sum_{\bf n}\langle (r^\alpha_{\nu{\bf n}})^2
                                                  \rangle_{T=0}
        =\sum_\lambda\left(\frac{{\rm e}^\alpha_\nu(\lambda)}
                               {\sqrt{m_\nu}}\right)^2
                          \frac{\hbar}{2\Omega_\lambda}.
\end{equation}
The resulting values are 
$\sqrt{\langle \overline{ (r^\alpha_\nu)^2} \rangle}=0.029$, $0.035$,
$0.048$, and $0.053$ {\AA} for $r^\alpha_\nu=r^z_\mCu$, $r^y_\mGe$,
$r^x_\mO$, and $r^y_\mO$, respectively. They are consistent with the
values of the total zero-point fluctuations obtained from the shell
model and the neutron-scattering experiments presented in
Ref.~\onlinecite{Braden98CGOanh}. The zero-point fluctuations are thus
a factor of 5 to 10 larger than the static distortions as given in
Eq.\ (\ref{rVector}).

On the other hand, the Ginzburg criterion discussed in Sec.\ 
\ref{sectionGinzburg} and the consistency of our results with
experimental ones justify our mean-field approach. In accordance with
that, Kl\"umper {\it et al.}\cite{Kluemper98CGO} show that a variety
of physical quantities can be obtained correctly in a mean-field
picture. It is beyond the scope of this paper but certainly an
interesting question addressed to future studies which quantities are
sensitive to the zero-point fluctuations and why.


\section{Discussion of $\chi_0$}
\label{sectiondiscussion}

The approach by Cross and Fisher\cite{CF79,Gros98CGO} gave a value of
$\chi_0\approx 0.26$. This value is independent of $J$ because of the
scale invariance at $q_c=\pi/c$. The scaling hypothesis is applicable
close to the critical point of the spin chain, i.e., in the limit
$T\to 0$. Recent DMRG results obtained by Kl\"umper {\it et
al.}\cite{Kluemper98CGO,Kluempertbp} show a strong temperature
dependence of $\chi_0(T/J)$. For $J_2=0$ and $J=120$ K they found
$\chi_0(T_{\rm SP}/J)\approx 0.28$. For $J_2/J=0.241$ and $J=150$ K
the parameter attains $\chi_0(T_{\rm SP}/J)\approx 0.56$, whereas for
$J_2/J=0.35$ and $J=160$ K they found $\chi_0(T_{\rm SP}/J)\approx 1$. 

The exact value of $J_2$ in CuGeO$_3$ has not yet been
determined. Fits to the susceptibility for $T>T_{\rm SP}$ indicate an
overcritical $J_2$,\cite{Riera95CGO,Fabricius98CGO} but fits to the
four-spinon continuum seen by Raman scattering\cite{Gros97CGO}
indicate an undercritical $J_2$. In favor of an undercritical $J_2$
is also the small binding energy of the singlet bound state for
$T<T_{\rm SP}$, as seen by Raman experiments.\cite{Sekine98} Interchain
coupling will reduce the value of $\chi_0$ because of an enhancement
of the antiferromagnetic correlations.\cite{Uhrig97CGO} 

As can be seen from Eq.\ (\ref{T_SP}) our coupling constants scale as
$g_\lambda \sim \sqrt{\chi_0}^{-1}$. From the above results follows
$1<\sqrt{\chi_0}^{-1}<1.9$ and we adapt the mean value of $\chi_0=0.5$
for our calculations. This value is close to the result for
$J_2/J=0.241$. Within the accuracy of our approach we can use $J=150$ K
as given by Castilla {\it et al.}\cite{Castilla95CGO} The choice of
$\chi_0$ is justified \`a posteriori by the agreement of the results
in the literature. Also note that including a NNN term with
$J_2/J=0.24$ in Eq.\ (\ref{Hdelta}) with $\delta_J=0.1$ and using
exact diagonalization gives a value of $\sum_{l_z} (-1)^{l_z} \langle
{\bf S}_{l_z}\cdot{\bf S}_{l_z+1} \rangle/N = 0.57$ per two Cu ions in
agreement with the value given in Eq.\ (\ref{S/N}). 

Applying hydrostatic pressure the transition temperature grows at a
rate of $4.8$ K/GPa.\cite{Takahashi95CGO} In our approach $T_{\rm SP}$ is
given by Eq.~(\ref{T_SP}) and depends on the coupling constants
$g_\lambda$, the frequencies $\Omega_\lambda$, and the factor
$\chi_0$. The coupling constants $g_\lambda$ in 
turn depend on the linear derivatives of the magnetic exchange
$g_\nu^\alpha$ and the polarization vectors, as given in
Eq.~(\ref{geff}). In a harmonic lattice the phonon frequencies and
polarization vectors are independent of pressure. It seems very
unlikely that the Peierls-active modes exhibit extremely large
negative Gr\"uneisen Parameters which would be needed in order to
describe the increase of $T_{\rm SP}$ upon pressure via the pressure
dependence of the phonon frequencies. The linear
coupling constants $g_\nu^\alpha$ also are independent of pressure,
and since the lattice distortions are rather
small,\cite{Braeuninger97CGO,Bradentbp} we do not expect higher-order
contributions to play a crucial role. We must thus conclude the value
of $\chi_0$ to be strongly pressure dependent.  

Together with the pressure dependence of $J_2/J$ discussed by
Fabricius {\it et al.},\cite{Fabricius98CGO} this may explain the
shift of $T_{\rm SP}$.\cite{Kluempertbp} When introducing interchain
coupling, prefactors and the functional dependence of the spin-spin
correlation function are also altered.\cite{Schulz86LL} The
compressibility of the crystal is largest in the $b$ direction so that
the alternation of the interchain coupling under pressure is another
possible origin of the pressure dependence of $\chi_0$ and $T_{\rm SP}$.


\section{Summary}\label{sectionsummary}

In this paper we have given a detailed analysis of the microscopic
magnetoelastic coupling in CuGeO$_3$ which may be easily transferred
to other systems. The comparison of several theoretical and
experimental approaches yields a satisfactory consistency. Numbers
have been given in Table \ref{summaryJofangle} for the angular
dependence of the magnetic exchange, in Table \ref{summarydelta} for
the dimerization, and in Table \ref{summaryJofP} for the pressure
dependence of the magnetic exchange. The quantitative agreement of
course is limited by the uncertainties within experiments and
theoretical techniques. Coupling constants for effective
one-dimensional real-space model Hamiltonians accessible to numerical
studies are given in Table \ref{tablereal}. We have discussed the
applicability of static models (Sec.\ \ref{sectiondimer} and
\ref{sectionrealspace}), and we were able to explain qualitatively the
$c$-axis anomaly of the thermal expansion (Sec.\
\ref{sectionthermal}).


\section*{Acknowledgments}

We are thankful to A.~Kl\"umper and R.~Raupach for giving us access to
results prior to publication and for discussions. 
We thank W.~Weber for discussions and for pointing out the
significance of the zero-point fluctuations. We acknowledge fruitful
discussions with B.~B\"uchner, H.~Fehske, S.~Feldkemper, T.~Lorenz,
U.~L\"ow, and W. Reichardt. The support of the DFG is greatfully
acknowledged.


\end{document}